\documentclass[fleqn,10pt]{wlscirep}
\usepackage[utf8]{inputenc}
\usepackage[T1]{fontenc}

\title{Rethinking infrastructure design: Evaluating pedestrians and VRUs' psychophysiological and behavioral responses to different roadway designs}

\author[1]{Xiang Guo}
\author[2]{Austin Angulo}
\author[3]{Arash Tavakoli}
\author[1]{Erin Robartes}
\author[1]{T. Donna Chen}
\author[1,*]{Arsalan Heydarian}

\affil[1]{Department of Engineering Systems and Environment, University of Virginia, Charlottesville, VA, 22904, USA}
\affil[2]{Department of Civil, Structural and Environmental Engineering, University at Buffalo, State University of New York, Buffalo, NY, 14260, USA}
\affil[3]{Department of Civil and Environmental Engineering, Stanford University, Stanford, CA, 94305, USA}

\affil[*]{ah6rx@virginia.edu}

\begin{abstract}
The integration of human-centric approaches has gained more attention recently due to more automated systems being introduced into our built environments (buildings, roads, vehicles, etc.), which requires a correct understanding of how humans perceive such systems and respond to them. This paper introduces an Immersive Virtual Environment-based method to evaluate the infrastructure design with psycho-physiological and behavioral responses from the vulnerable road users, especially for pedestrians. A case study of pedestrian mid-block crossings with three crossing infrastructure designs (painted crosswalk, crosswalk with flashing beacons, and a smartphone app for connected vehicles) are tested. Results from 51 participants indicate there are differences between the subjective and objective measurement. A higher subjective safety rating is reported for the flashing beacon design, while the psychophysiological and behavioral data indicate that the flashing beacon and smartphone app are similar in terms of crossing behaviors, eye tracking measurements, and heart rate. In addition, the smartphone app scenario appears to have a lower stress level as indicated by eye tracking data, although many participants don't have prior experience with it. Suggestions are made for the implementation of new technologies, which can increase public acceptance of new technologies and pedestrian safety in the future. 
\end{abstract}
\begin{document}

\flushbottom
\maketitle

\section*{Introduction}

At its core, infrastructures are in fact an engineering product that have significant impact on people’s day to day lives. However, unlike many other products, we often overlook the importance of changing the design based on user feedback within the design phase. This is partly due to the fact that such a process can become costly and often not practical in the context of designing infrastructures at the community and city scales \cite{naumann2011design}. For instance, it is not possible to build different replicas of the same road for testing driver distraction in each alternative design of the road. As a result, many times, design features are chosen by the designer and engineers with minimal feedback (if any) from all end-users (e.g., drivers, bicyclists, pedestrians, scooterstis that will use the road in the future).

Over the recent years, due to advancements in technology, designers have started to take into account the end user and human factors especially in the areas of human-building interaction, and human-vehicle interaction. This approach, which is often referred to as a human-centric approach in design, tends to put the user's needs, comfort levels, and preferences at the center of the design process \cite{amundadottir2017human,li2021human}. The integration of human-centric approaches has gained more attention recently due to the fact that more automated systems are being introduced into our current built environment, which requires a correct understanding of how end-users perceive such systems and respond to them. Additionally as the post-construction cost of changing the design is often very high and impractical, it is important to be able to test different infrastructural design alternatives and receive end-user feedback before the construction phase.

For the design to become human-centered, it is crucial to measure end-user feedback, feelings, perceptions of comfort and safety within different alternative designs. To measure human response and feedback in these systems, a variety of factors and information are needed to be considered. This requires methods that can quantitatively provide insights on the perception of the infrastructure from different end-user perspectives. Thus, it is required to have a platform to (1) simulate the design, and (2) assess users' feedback, feelings, perceptions of comfort and safety while using the design objectively within the simulated platform. As a new emerging technology, Immersive virtual environments (IVEs) simulators are a promising tool for behavioral studies and to identify how end-users perceive and react to different design alternatives. Within IVEs, users have the ability to realistically visualize and interact with the infrastructure before construction. IVE has been applied in many indoor human-building interactions, such as buildings \cite{heydarian2015immersive,francisco2018occupant} and driving \cite{lee1998driving,sportillo2018get,chung2022static}. 

Coupling IVEs with psycho-physiological sensing of users allows researchers to measure the effect of infrastructure design on the end-user perception and response, which can objectively prevent faulty design features prior to construction. Human psycho-physiological metrics such as heart rate, skin temperature, skin conductance, and eye gaze patterns were used in literature for assessing human state such as stress level, emotion, and cognitive load \cite{lohani2019review,kim2018stress, tavakoli2021harmony,chesnut2021stress, tavakoli2022multimodal}. Further, these physiological measures have also shown to be more sensitive than task performance measures to identify task difficulty when using a new technology or exploring a new environment\cite{ikehara2005assessing}. While experiencing certain states, human biomarkers undergo certain changes that may be linked to these states as a coping mechanism and by monitoring these biomarkers, we can detect these states automatically \cite{chesnut2021stress}. 

Within infrastructural elements, designing proper roadway and transportation systems is of high importance as it is highly associated with users well-being, injuries, fatalities, and overall quality of life. Roads play a major role in how citizens are mobilized across different built environments and if designed properly, can play a major role in improving health and well-being of citizens. However, there is very limited attention to how roadway systems need to be designed to be inclusive for all users. The majority of research on roadway design has heavily focused on studies evaluating driver's behavior, safety, and responses to different design conditions and contextual settings \cite{horberry2006driver}. As a result limited studies have focused on other road users such as pedestrians, bicyclist's responses to different roadway design and conditions.

Among all road users, Vulnerable Road Users (VRUs) such as pedestrians and bicyclists are gaining more attention as they have an increasing fatalities in recent years \cite{national2020overview,world2018global}. Among these, VRUs, pedestrians are facing more safety challenges on the road, especially during the midwalk crossing as they have less protective equipment and slower speed than the vehicles, scooters, and bicycles \cite{tezcan2019pedestrian}.  National Highway Traffic Safety Administration (NHTSA) has reported a 35\% increase in pedestrian fatalities nationwide between 2008 and 2017 \cite{national2020overview}. In addition, nearly 300,000 pedestrians are killed on roads globally each year, accounting for 22\% of overall fatalities of road accidents \cite{world2018global}. These trends indicate that the design of current roadways needs to be improved for all users, especially for the vulnerable road users like pedestrians.  

Ensuring the safety of pedestrians is a challenge for researchers, as pedestrian's decision to cross and the crossing behavior may be affected by many factors, such as pedestrian infrastructure, roadway design, traffic volumes, vehicle speed, and visibility of the road environment \cite{stoker2015pedestrian,cloutier2017outta}. Accidents involving pedestrians are especially common at un-signalized and mid-block crosswalks, where vehicles are less likely to yield to pedestrians \cite{markkula2022explaining}.

To increase pedestrian safety at mid-block crossings, different safety treatments that utilize new technologies and designs have been introduced. For instance, rapid flashing beacon (RFB) is a common treatment at mid-block crossings \cite{fitzpatrick2015investigating}. When a pedestrian intends to cross the street, after pressing the button, the rapid flashing beacon can alert nearby drivers about the intention to cross the street so the drivers can yield in advance. RFBs have been shown to increase yield rate (86\%) especially when the speed limit is higher than 35 miles per hour \cite{fitzpatrick2014driver}. In addition, a driver is more than three times as likely to yield when a beacon has been activated as when it has not been activated \cite{fitzpatrick2015investigating}. Other efforts such as pedestrians level of control have been also studied for midblock crossings. Specifically, in a simulated two-way traffic environment study, a vibrotactile wristband is designed to help older pedestrians make safer street-crossing decisions. Although the percentage of decisions that led to collisions with approaching cars decreased significantly, only 51.6\% of the time did subjects respond in accordance with the wristband \cite{coeugnet2017vibrotactile}. Countdown timer is another widely used treatment, however, since the pedestrian often overestimate their speed, there is a higher chance of red light running as they choose to start crossing even if the remaining time is short. A study in 2018 implemented new timers with required pedestrian crossing speeds to help their judgments based on remaining time; however the results indicated that they could not eliminate the chance of running into red light altogether \cite{zhuang2018display}. Pedestrian footbridge is able to separate the pedestrians from the traffic flow, but it usually implies longer walking distances compared to the direct crossing. In \cite{cantillo2015modelling}'s study, a hybrid choice model shows longer walking distance to safer options increases probability of direct crossing, and pedestrians prefer cross through a signalized crosswalk than to a footbridge. Another study also shows that footbridges and underpasses were systematically rated below signalised crossings \cite{anciaes2018estimating}. The development of connected vehicles and autonomous vehicles has changed the communication environment between pedestrians and vehicles. Recent studies have focused on how to communicate awareness and intent of autonomous vehicles to pedestrians \cite{mahadevan2018communicating}. However, very few studies have pedestrian-centered design, which is how to communicate the pedestrian's crossing intentions to the vehicle. Similar to building environment user studies, IVEs can provide us with the opportunity to evaluate new technologies and modes of communications between the pedestrian to vehicles to better understand the comfort of pedestrians and usability of these new technologies. However, limited studies have utilized IVEs for such pedestrian-centered design \cite{velasco2019studying}. 

Within the pedestrian safety research, studies have integrated the aformentioned physiological signals to examine participants' perceived safety, and comfort within different urban environment. The studies in this area which goes as far back to 1971, showed that blind pedestrians experienced higher heart rate when walking unaided \cite{peake1971use}. Kitabayashi et al. used heart rate as the biosignal and found out that pedestrians stress in walking are affected by the road congestion \cite{kitabayashi2015analysis}. Additionally, pedestrians' physiological measures were also shown significantly correlated with certain urban features such as uneven sidewalks as well as subjective ratings of walkability \cite{kim2019influence,kim2022capturing}. Thus, combining physiological metrics within the IVE can help us better understand the effect of each pedestrian-centered design on road users.

This paper aims at assessing different pedestrian crossing designs features that are pedestrian-cenetered by informing the incoming vehicles on the intend of the pedestrian. The research objectives are: 1.identify the benefits and limitations of using IVEs for collecting and modeling VRUs’ behaviors and psycho-physiological responses while highlighting how such information could improve the design decision making. 2. Evaluate the objective and subjective measures of perceived safety rating across different alternative designs. 3. Evaluate pedestrians' crossing behavior and psychophysiological responses across different conditions.  

We introduce a data-driven decision-making methodology to evaluate suitability, acceptance, and use of innovative technologies for alternative infrastructure designs. We first introduce the system framework to collect VRUs’ behavior and physiological response. The system has integrated data collection methods (pedaling/walking performance, eye tracking, heart rate, and video) in virtual reality, and the modularized components makes it applicable to evaluate infrastructure design for all roadway users whether they are pedestrians, bicyclists, scooterists, construction workers, or drivers. Through a case study of pedestrian crossing, 51 pedestrians' stated preferences, crossing behaviors and physiological responses are collected and analysed with three different mid-block crossing safety treatments – painted crosswalk (as-built), rapid flashing beacons (flashing beacon, Figure \ref{fig:RFB}), and a connected vehicle phone application (smartphone app, Figure \ref{fig:phoneapp}).

This study hypothesises that:

\begin{itemize}
    \item H1: Pedestrians would subjectively feel less safer in the as-built scenario, and the flashing beacon scenario would be perceived as the safest design compared to the other two alternatives.
    \item H2: Pedestrian would have safer crossing behaviors (longer wait time, reduced crossing time with less stops), when flashing beacons or a smart app is available to communicate with the drivers. 
    \item H3: Pedestrian in the flashing beacon and smartphone app conditions will have a lower cognitive workload, as indicated by lower mean fixation length, higher fixation rate, lower gaze entropy and lower mean heart rate.
    \item H4: Pedestrian would not provide significant differences in  psychophysiological and behavioral responses in the flashing beacon and smartphone app conditions.
\end{itemize}

\section*{Results}
\subsection*{Stated Preference Survey Response}
On average, participants have a higher safety rating in flashing beacon scenario (4.56 out of 5 scale), followed by the smartphone app (3.6), and the as-built environment (3.0). The differences between the safety rating are all significant at a 95\% confidence level ($\alpha = 0.05$).  

\begin{figure} 
    \centering
    \includegraphics[width=\linewidth]{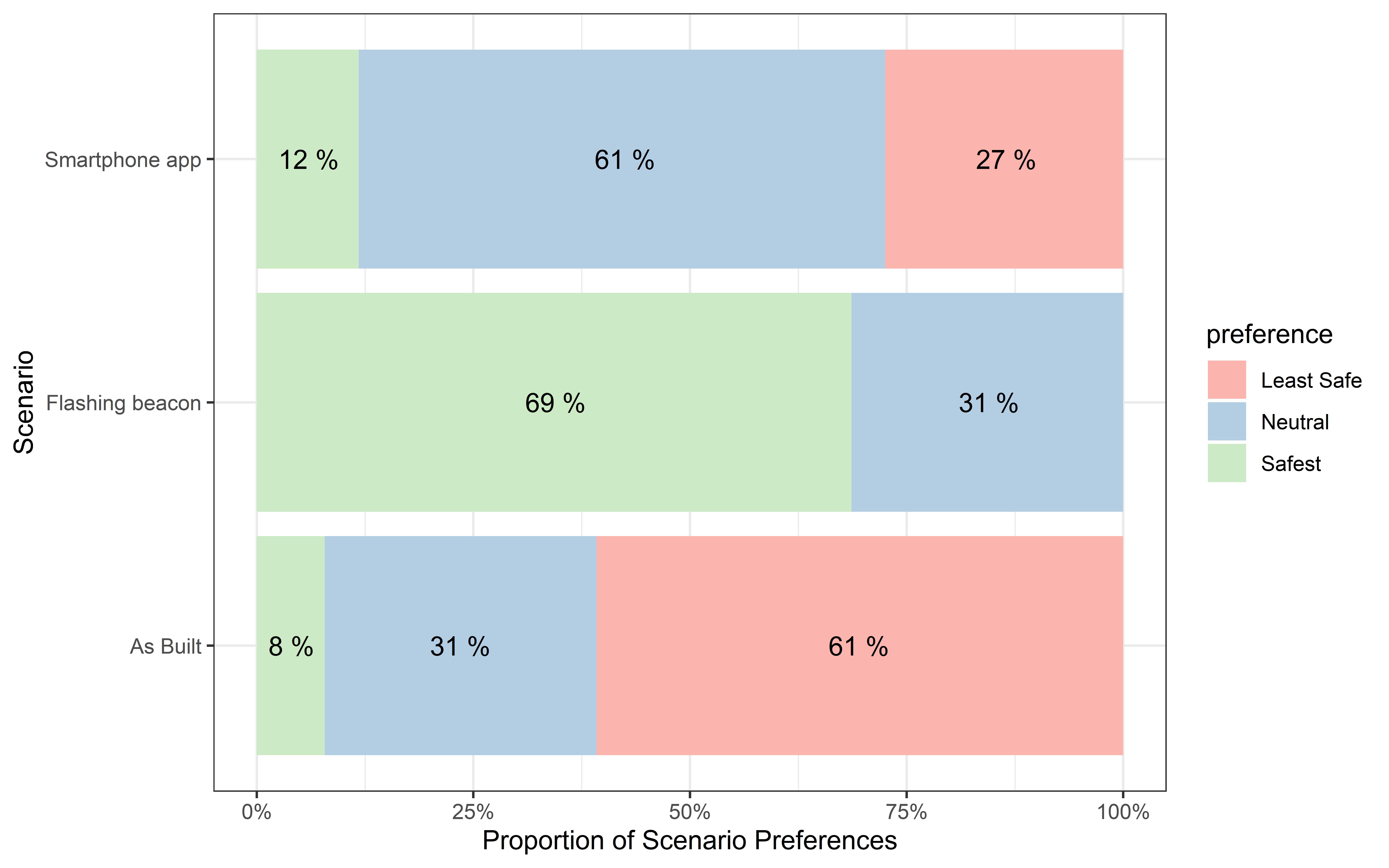}
    \caption{Safety preference for different scenarios from survey response}
    \label{fig:safety_preference}
\end{figure}

Additionally, once asked to rank the three environments based on their perceived safety measures from safest to least safe, participants' responses supported the previous metric, with flashing beacon ranked as the safest and the as built environment as the least safe condition (Figure \ref{fig:safety_preference}). 69\% of the participants rate the flashing beacon scenarios as the safest scenario of the three options, none of them rate it as the least safe scenario. For smartphone app scenario, 12\% rate it as the safest, 27\% participants rate it as the least safe one. For the as-built scenario, only 8\% rate it as the safest and 61\% of the participants rate it as the least safe scenario. The results from the stated preference survey response support the H1.

\subsection*{Crossing Behavior}

\subsubsection*{Crossing Time}
\begin{figure} 
    \centering
    \includegraphics[width=\linewidth]{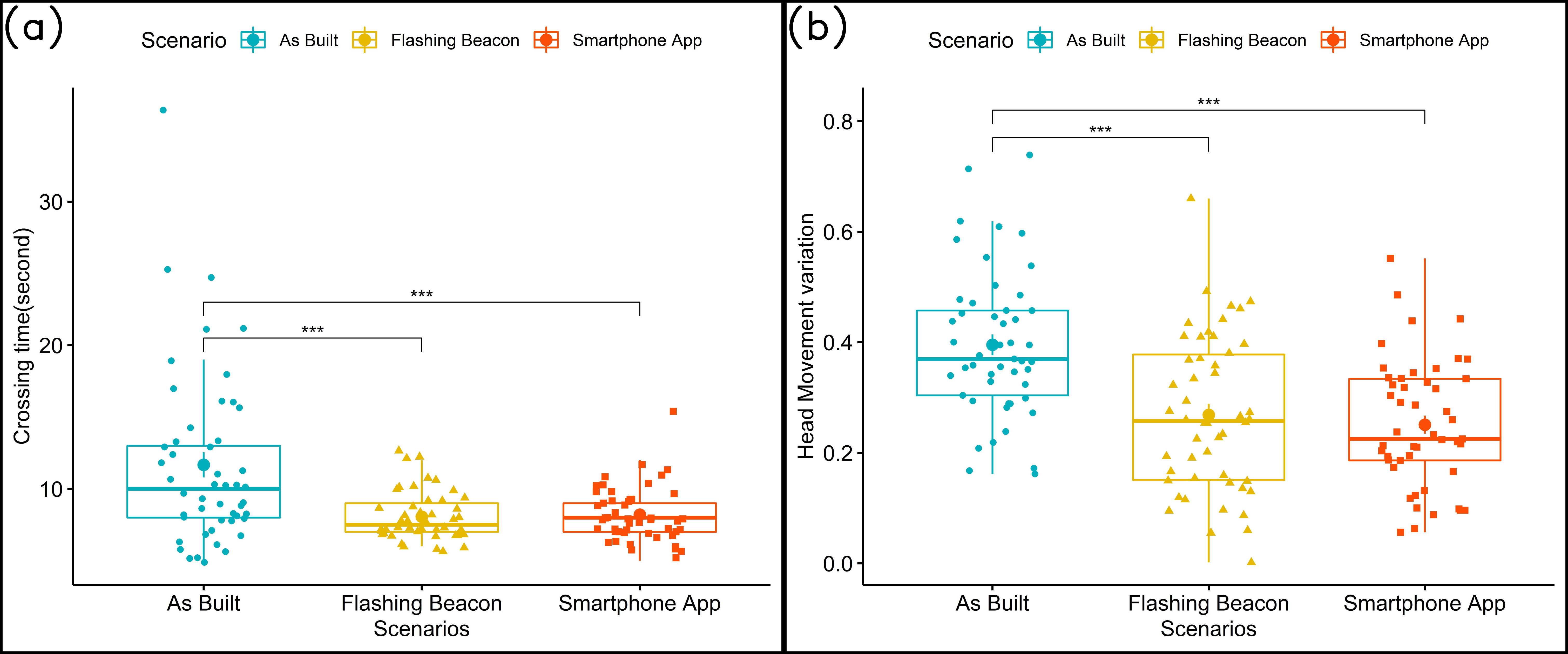}
    \caption{(a) Average crossing time of different scenarios; (b) Head movement variation of different scenarios}
    \label{fig:crossing_behavior}
\end{figure}

For the crossing time, as shown on Figure \ref{fig:crossing_behavior}(a), participants had a significantly lower crossing time in the flashing beacon ($\beta = -3.604, SE = 0.717, p < 0.0001$)  and smartphone app cases ($\beta = -3.417, SE = 0.720, p < 0.0001$) as compared to the as-built environment. No significant differences are found between the flashing beacon and smartphone scenarios.

\subsubsection*{Wait time before crossing}
The wait time before crossing for as-built, flashing beacon and smartphone app scenarios are 20.34s, 22.20s, and 21.47s respectively. A significant difference between as-built and smartphone app scenarios is found ($\beta = 1.82, SE = 0.72, p < 0.05$).

\subsubsection*{Wait time after crossing decision}
The wait time after crossing decision for smartphone app scenario (mean = 4.23s, sd = 2.92s) is lower than flashing beacon scenario (mean = 5.20s, sd = 3.35s), but the difference is not significant.

\subsubsection*{Head Movement}
The result shows a significant difference between the as-built and the two other scenarios with both p-values less than 0.001. Specifically, for the flashing beacon scenario,$\beta = -0.128, SE = 0.0227, p < 0.0001$, and for the smartphone app scenario, $\beta = -0.145, SE = 0.0227, p < 0.0001$. However, we did not find a difference between the flashing beacon and smartphone app scenario. As shown on Figure \ref{fig:crossing_behavior}(b), participants had a higher variation of head movement direction in the as-built environment as compared to the other two scenarios.  

\subsubsection*{Stop during crossing}
We manually annotated the experiment videos to determine if the pedestrians have stopped in the middle of their crossing. Two participants' (participants 42 and 46) data are excluded due to failure in video recording. Therefore, 49 participants' stop behaviors are recorded. As shown in Table \ref{table:pedestrian stop crossing}, Pedestrians in the as-built scenario stop significantly more in the middle of the corss walk comapared to the other two scenarios. Interestingly, the flashing beacon and smartphone app scenarios has exact the same number of stops across the participants. In both scenarios, there are 10 participants who stop in the middle of crossing to wait for the vehicle's response although they are told that the vehicles will stop for them after they send their request by pushing the buttons.

\begin{table}[h!]
\caption{Number of pedestrians' stops during crossing}
\label{table:pedestrian stop crossing}
\centering
 \begin{tabular*}{0.45\textwidth}{c c c} 
  \hline
 Scenarios & No stop cases & Stop cases \\ [0.5ex]  \hline 
 As-built & 19  &  30 \\
 Flashing beacon & 39 & 10 \\
 Smartphone app & 39 & 10 \\
 \hline
 \end{tabular*}
\end{table}

The crossing behavior results are aligned with H2.

\subsection*{Eye Tracking}
For the eye tracking data, five participants' data are excluded due to hardware failure during data collection. The eye tracking results in this section are based on 46 participants' data.

\subsubsection*{Fixation}
\begin{figure} 
    \centering
    \includegraphics[width=\linewidth]{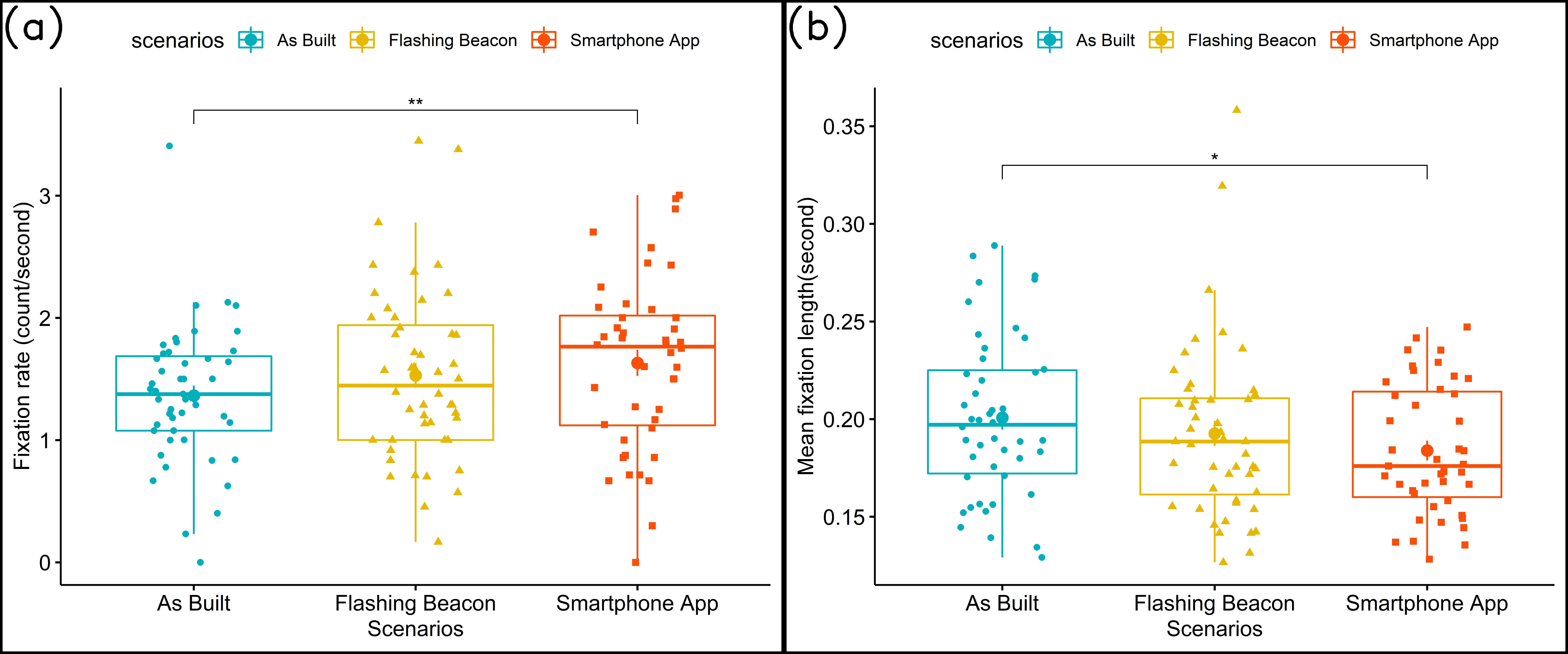}
    \caption{Fixation measurement during crossing of three scenarios, (a) Fixation rate, (b) Mean fixation duration}
    \label{fig:fixation}
\end{figure}

Participants in the smartphone app scenario had a significantly higher fixation rate as compared to the as-built environment ($\beta = 0.270, SE = 0.100, p < 0.01$). We did not find any significant differences between the other scenarios, as shown in Figure \ref{fig:fixation}(a).

For mean fixation duration, there is a significant difference between the as-built and the smartphone app scenarios ($\beta = -0.0179, SE = 0.00787, p < 0.05$). As shown on Figure \ref{fig:fixation}(b), participants had a lower mean fixation duration in the smartphone app scenario with an average of 0.184 seconds as compared to the as-built environment with an average of 0.201 seconds, in between is the mean fixation duration of flashing beacon scenario (0.193 seconds). 

\subsubsection*{Gaze Entropy}
\begin{figure} 
    \centering
    \includegraphics[width=\linewidth]{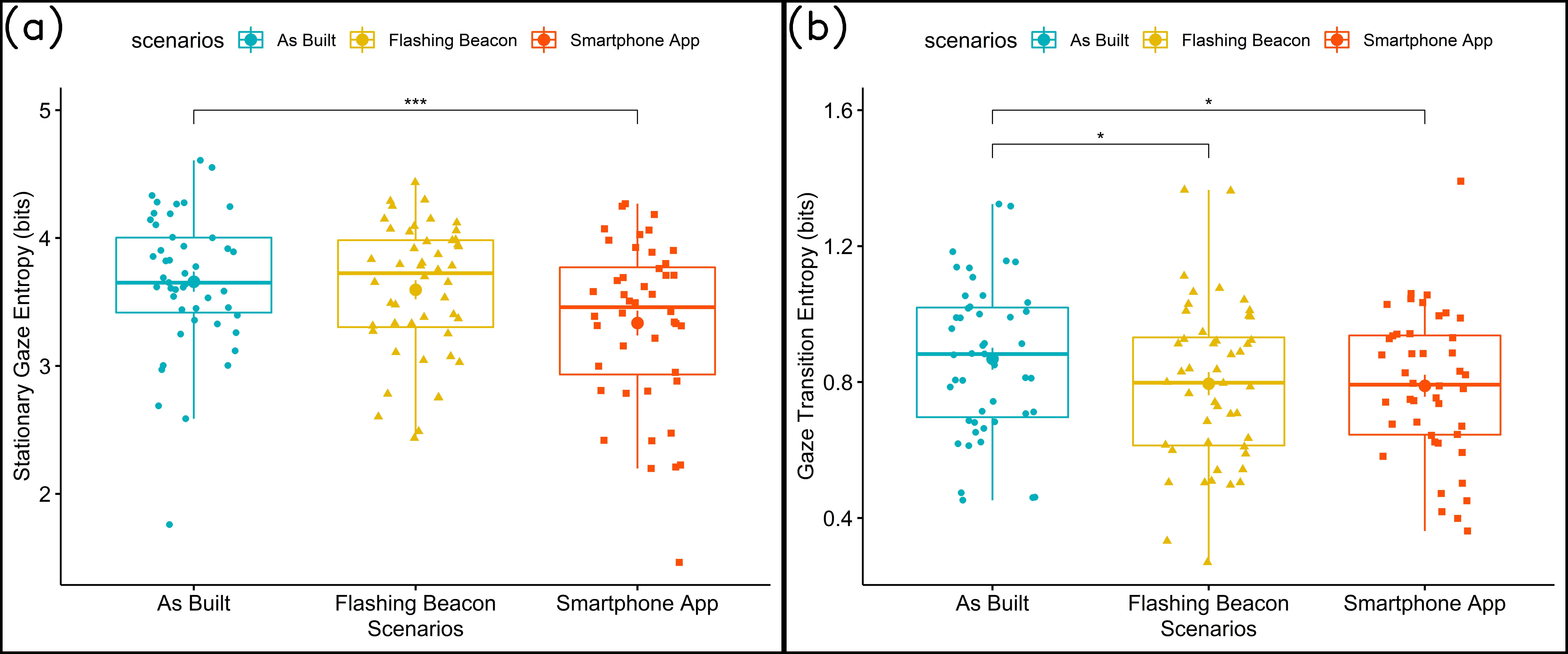}
    \caption{Gaze entropy during crossing of three scenarios, (a) Stationary gaze entropy, (b) Gaze transition entropy}
    \label{fig:Gaze_entropy}
\end{figure}

There are two types of gaze entropy measures: stationary gaze entropy (SGE) and gaze transition entropy (GTE). The results for the SGE shows that participants had a significantly lower SGE in the smartphone app as compared to the as-built environment ($\beta = -0.0830, SE =  0.0411, p < 0.05$). However, No significant difference among the other scenarios are found. The results of GTE shows that the GTE is significantly lower in the both flashing beacon ($\beta = -0.0764, SE = 0.0403, p < 0.05$) and smartphone app scenarios ($\beta = -0.0830, SE = 0.0411, p < 0.05$) as compared to the as-built environment. No significant results are found between the flashing beacon and smartphone app scenario, as shown in Figure \ref{fig:Gaze_entropy}(b).

\subsection*{Heart Rate}
The heart rate result indicates that there is no significant differences between the three scenarios a 95\% confidence level. Marginal difference for the mean heart rate during crossing is found between the smartphone app scenario and the as-built scenario ($\beta = -1.909, SE = 1.109, p = 0.0886$). The mean HR (beat per minute) of the as-built, flashing beacon and smartphone app are 86.40, 86.29 and 84.63, respectively. The results of eye tracking and heart rate support the H3 and H4. 

\section*{Discussion}
Overall, from stated preference results, both the flashing beacon and smartphone app scenarios are perceived to be safer than the as-built scenario, and the participants show a higher preference on the flashing beacon scenario based on both the subjective and objective ratings. The majority of the participants (69\%) choose the flashing beacon as the safest scenario, which could imply their trust in this technology (as well as their familiarity with this technology as it exists on some roads).

Interestingly, the results from crossing behavior and physiological responses are slightly different from stated preference. For average crossing time, both the flashing beacon and smartphone app scenarios have a lower average crossing time compared to the as built scenario; additionally, there is no significant differences between the flashing beacon and smartphone app crossing time. The pedestrians have a lower wait time before crossing, but spend more time during the crossing, this is aligned with an observational study conducted by \cite{tezcan2019pedestrian} at midblock crosswalks in which pedestrians who waited for little or no at the curbside generally lost time during the crossing.

It is important to also note that some participants indicated that they were not sure about the smartphone app performance, so they chose to wait until the vehicle came to a complete stop for them. However, when checking the waiting time after crossing decision, the smartphone app scenario actually has a lower average waiting time after crossing decision (4.23s) than flashing beacon scenario (5.20s), although the difference is not significant. This  may be explained by different reactions required by the two interactions (press the button on the phone vs. physically reach out to the button), or the gap acceptance difference in the two scenarios. 

With respect to head movement, the larger head movement variation in the as-built scenario indicates that participants are more hesitant during crossing, while no significant differences are found between the two alternative designs. Furthermore, visual inspection of the videos also qualitatively verifies the fact that the proportion of stop behaviors during crossing are the same for the two alternative designs, and both are lower than the as-built scenario. 

For eye tracking data, the difference in fixation rate and mean fixation duration between the as-built and smartphone app scenario shows pedestrians' different visual scanning strategies. The longer fixation duration in the as-built scenario means that pedestrians spent a long time on searching the environment and potential hazards. As reported by previous studies, longer fixation duration and lower fixation rate is related to higher cognitive load \cite{liu2022assessing}. An earlier pedestrian eye tracking study also found that 'safe' pedestrians have a lower mean fixation duration than 'rogue' pedestrians after they get used to the environment \cite{jovancevic2009adaptive}. Lower SGE and GTE are observed in the smartphone app, as far as we know, there is no existing studies about the pedestrian gaze entropy. In flight situations, low gaze entropy is usually accompanied by high situation awareness, for different tasks, the gaze entropy of the group that succeeded in the task was low \cite{bhavsar2017quantifying}. Therefore, our results may indicate that the smartphone app scenario may have a lower cognitive workload for the pedestrian to cross.  

Due to the relatively low HR data frequency, only a limited number of HR data points are utilized for the mean HR comparison. Marginal significantly lower mean HR is found for smartphone app scenario in this study, which may reveal a lower stress level in the smartphone app scenario as compared to the other two scenarios. As mentioned before, previous studies show that lower HR values are generally associated to calmer, less-stressful states \cite{kim2018stress}. However, we note that this finding needs to be validated in the future study with more professional HR data collection devices.

The qualitative feedback collected from participants may also help to find the reasons behind the differences in subjective ratings and objective responses. A couple of participants stated that they were not sure about what would happen in the smartphone app scenario after pressing the button on the screen although instructions are given before the experiment. This may be the reason why more participants prefer the flashing beacon scenario. However, the crossing behavior data shows that the waiting time after crossing decision for the smartphone app is not significantly different from flashing beacon. For other crossing behavior variables, we also do not find significant differences between the flashing beacon and smartphone app scenario. In addition, for physiological responses, the smartphone app scenario seems to have a slightly better overall performance with a shorter fixation duration, higher fixation rate and lower HR, which is related to lower cognitive load. Given the fact that there is still much room for improvement in the smartphone app scenario, a better physiological performance can be expected if such limitations are addressed. 

Our results further emphasize the importance of objective measurement for the evaluation of infrastructure designs as the users' subjective answers may not reflect their actual behaviors. The difference in subjective ratings and objective responses also highlights that public education is an important step of new technology implementation. In our study, although the smartphone app scenario shows a good overall performance, participants don't have a high safety ratings on it because they don't have any related experience with the new technology. IVE-based simulation offers a risk-free and low-cost platform for the public to get familiar with new technologies, which will help to increase the acceptance of these new technologies which are currently not familiar to them.

\section*{Limitations and Future Work}
The eye tracking section of our study only focuses on the overall information of fixations (fixations rate and mean fixation duration) and the general distribution of the fixations (gaze entropy), it makes more sense to extract contextual information about fixations. By defining Area of Interests (AOIs) such as the button on the flashing beacon, smartphone, crosswalk path or other vehicles, it would help to gain a better understanding of what the pedestrians are looking at. The visual attention allocation of pedestrians will provide more information about distraction state \cite{gruden2021pedestrian}. In our future study, in-depth analysis of eye tracking data by integrating the AOIs information will be performed to explore pedestrians' visual attention allocation on key AOIs, such as the flashing beacon button, the smartphone and the vehicles.

Another limitation of our study was the low frequency of HR data. Due to collecting HR using off-the-shelf smartwatches we did not have access to higher frequency physiological sensing. Future work should consider adding other physiological sensing modalities such as skin temperature and skin conductance to enhance the physiological sensing module and inference. However, it should also be considered that more devices might degrade the feeling of realism of the study. More advanced devices that can collect multiple physiological sensors simultaneously can be integrated into studies as such to keep the realism while recording a higher number of modalities of data. 

Other limitations, as also mentioned by the participants were to include (1) realistic vehicle actions, (2) feedback from the smartphone app, and (3) traffic simulation. Currently, we are improving the logic of the vehicle by refactoring the vehicle speed controller so the response will be more realistic. More ways of interactions and the feedback are being developed such as audio warning, tactile feedback from the controller, vehicle's flashing light, projections on the crosswalk, and so on. The various ways of interactions will be evaluated by users' stated preferences and objective responses as well. Moreover, based on our framework \cite{Guo2022}, it is possible to include multiple agents in the IVE, so other road users such as bicyclists and drivers can be studied together with pedestrians within the IVE. 

\section*{Conclusion}
This paper presents the evaluation of three pedestrian crossing infrastructure designs (the as-built painted crosswalks, the flashing beacon and a connected vehicle phone application) in an IVE-based experiment. With the system framework, the stated preferences, crossing behavior and physiological responses are collected from 51 participants. The results indicate that the two alternative designs have a higher safety ratings than the as-built scenario, and the flashing beacon scenario is rated as the safest. Pedestrians in the as-built scenario have a lower waiting time but spend/lost more time during crossing by stopping in the middle of the crosswalk to wait for the vehicle, in addition, a larger head movement variation is observed in the as-built scenario. The crossing behavior in the flashing beacon and smartphone app scenario is similar. For the eye tracking data, pedestrians had a shorter fixation duration, larger fixation rate, smaller stationary gaze entropy and smaller gaze transition entropy in the smartphone app than the as-built scenario, which may be resulted from a lower cognitive workload. The difference between the flashing beacon and as-built scenario is not as significant as the smartphone app. A marginal significant lower mean heart rate is found in the smartphone app scenario. Overall, both the flashing beacon and smartphone app have a better physiological performance than the as-built scenario, but the smartphone app scenario appears to have a slightly better physiological outcome. Qualitative feedback is collected from the participants to explore the reasons for the differences between stated preferences and objective measurements, discussions, and suggestions are made. In conclusion, public education is required before the implementation of new technologies such as connected vehicles, which can help to increase users' acceptance and safety.

\section*{Methods}
\subsection*{Study Design}
This research designs a within-subject experiment to study pedestrians' stated preferences, crossing behavior, and physiological responses to three different mid-walk crossing designs in an immersive virtual environment with a random order: painted crosswalk (as-built), rapid flashing beacons (flashing beacon), and a connected vehicle smartphone application (smartphone app). The selected location for this study is the intersection of Water St and 1st Street South in Charlottesville, Virginia. This place has been identified as a hotspot for pedestrian-vehicle accidents in the Virginia Department of Transportation's Pedestrian Safety Action Plan \cite{cole2018pedestrian}. The intersection of Water Street and 1st Street South is chosen as the study site. The north side of the intersection is a dead-end road (utilized only for deliveries). The south side of the road is a one-way street, which vehicles cannot turn onto from Water Street. 

In all the three scenarios, pedestrians will be placed into the beginning location, facing the crosswalk heading southbound along 1st Street, crossing Water Street from the north side of the road. The independent variables are the crossing infrastructure designs and demographic information (i.e., age, gender). The dependent variables are stated preferences of the three scenarios, crossing behavior (crossing time, waiting time before crossing, waiting time after crossing decision, stop or not during crossing, and head movement variation) and physiological responses (i.e., eye tracking and HR features) during crossing.

\subsection*{Virtual Reality System Setup}
A one-to-one road environment is built in the Unity software with SteamVR platform. HTC Vive Pro Eye headsets with the controllers are utilized for any interactions in the IVE. More detailed information of the IVE setup is available in our previous studies \cite{Guo2022,guo2021benchmarking,angulo4055270validation}. Vehicle traffic within the IVEs is generated from empirical gap acceptance data observed at the real-world location. The gaps between vehicles are generated to fit the empirical distribution of accepted gap sizes \cite{angulo4055270validation}. These gaps are randomized before each scenario so each participant's exposure to any gap is randomized. All the vehicles has a speed of 25 mph, followed the speed limit. Vehicle type is also randomized from the four vehicle models used in the IVE. 

\subsubsection*{As-Built scenario}
The as-built environment is built to model the existing painted crosswalk along the Water Street corridor to serve as the base case against the other two alternative designs. In the IVE, the pedestrian's task is to crossing the street when they feel safe to do so after the first vehicle passes the crosswalk. The vehicle will stop right before the crosswalk to wait for the pedestrian to cross if a conflict is expected to happen.   

\subsubsection*{Flashing beacon scenario}
\begin{figure} 
    \centering
    \includegraphics[width=\linewidth]{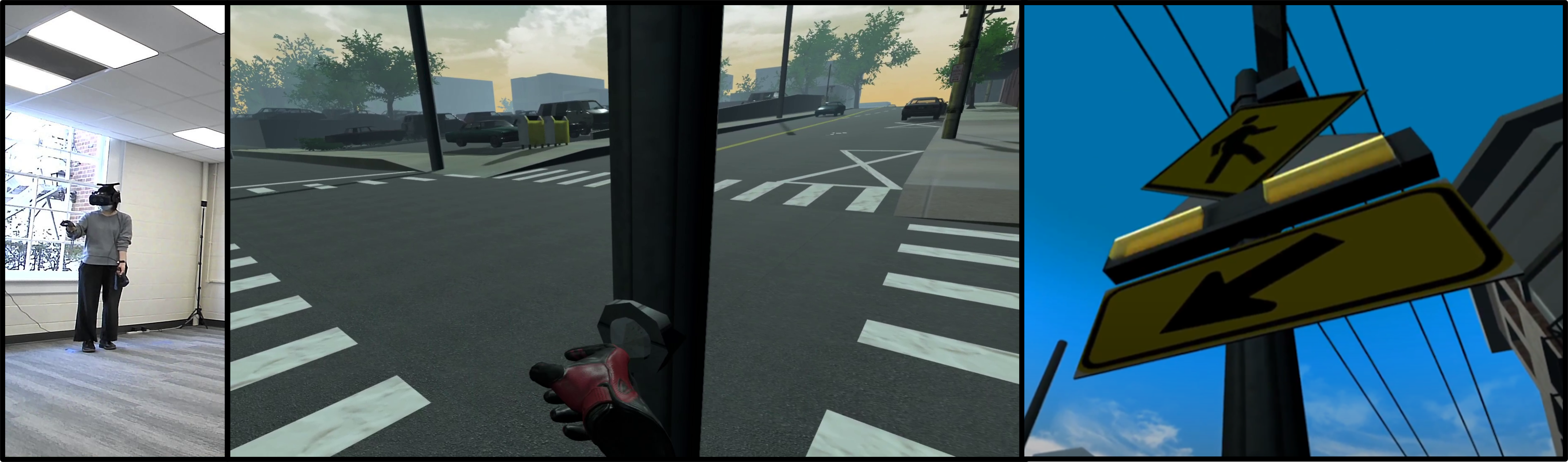}
    \caption{A pedestrian is using a flashing beacon for crossing (left), the pedestrian point of view in IVE (middle), the flashing beacon in IVE (right). a sample video is available in the following link \url{https://youtu.be/hz64mFP83LA}}
    \label{fig:RFB}
\end{figure}

In the flashing beacon scenario, the pedestrian is allowed to cross the road whenever they feel appropriate. Pedestrians are able to interact with the flashing beacon by pressing the button located on the sign pole to initiate the flashers on the beacon. Figure \ref{fig:RFB} shows how a pedestrian interacts with the RFB while in VR prior to crossing, as well as an image of the RFB in VR when used. 

\subsubsection*{Smartphone application}
\begin{figure} 
    \centering
    \includegraphics[width=\linewidth]{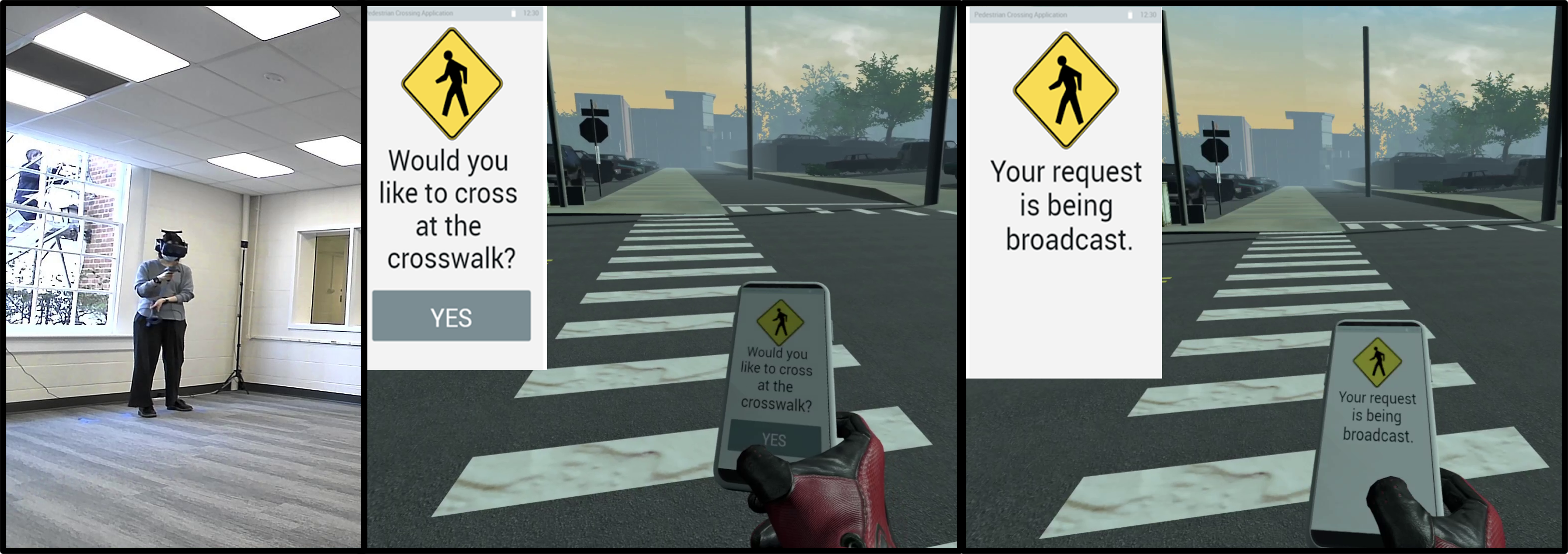}
    \caption{A pedestrian is using smartphone app (left), pedestrian's point of view in IVE (middle), smartphone app user interface before and after pressing (right), a sample video is available in the following link \url{https://youtu.be/Q_LUoIZuPKs} }
    \label{fig:phoneapp}
\end{figure}

In the smartphone app scenario, pedestrians will have a cellphone (a controller in their right hand in real life) in their right hand once they are placed in the IVE. As shown in Figure \ref{fig:phoneapp}, there are two interfaces that will show up on the phone during testing. The first interface of the mobile phone application (initial state) asks the pedestrian if they wish to cross the crosswalk. Should the pedestrian answer “Yes” and press the button on the controller's central pad, a new interface will pop up indicating “Your request is being broadcast”. The pedestrian is then free to cross the crosswalk and vehicles will yield before the crosswalk for the pedestrian.

\subsection*{Data Collection}
The data collection method of this study follows the framework of our previous study \cite{Guo2022}, different types of behavioral and physiological data are collected: stated preferences from pre and post experiment survey, crossing behavior data from Unity, eye tracking data from Tobii Pro Eye headset, heart rate data from smartwatches.

\subsubsection*{Survey Response}
In addition to demographic information, the participants are asked for their safety ratings and preferences after the experiment. For each scenario, they will be asked to provide their answer with a Likert Scale 1-5 to the question "How safe do you feel in the scenario", where 1 indicates "not safe at all" and 5 indicates "very safe". Furthermore, they are asked to rank the safest to the least safe scenario from the three environments.

\subsubsection*{Crossing Behavior}
Five response variables are recorded to represent the pedestrians' crossing behavior: crossing time, waiting time before crossing, waiting time after crossing decision, stop or not during crossing, and head movement variation. The crossing time is defined as the time interval from the moment when the pedestrian start crossing to the moment when the pedestrian reaches the other side of the crosswalk. Waiting time before crossing is defined as the time between the start of the experiment and the moment when the pedestrian start crossing. Waiting time after crossing decision are defined as the waiting time after pedestrian's decision to cross the street (after pressing the button either on flashing beacon or smartphone to start crossing), which is only accessible in the flashing beacon and smartphone app scenarios. Stop or not during crossing is a binary response about whether the pedestrian has a obvious stop to wait for the vehicle's behavior during crossing. The head movement is defined as the variations in the 3-D head movement direction in the unit vector.

\subsubsection*{Fixation}
Fixation is defined as the moments when eyes stop scanning about the scene and hold the central foveal vision in certain places to look for detailed information of the target object. Similar to previous studies \cite{shiferaw2019review,guo2022roadway,guo2022roadway}, We define a fixation with 25 ms minimum duration and 100 pixel maximum dispersion thresholds to extract the fixation information from the original eye tracking data and videos. Two measurements of fixation are calculated: (1) the mean fixation duration is defined as the average length of all fixation events during the crossing; and (2) the fixation rate is defined as the number of fixations per second during the crossing.

\subsubsection*{Gaze Entropy}
there are two types of gaze entropy measures: stationary gaze entropy (SGE) and gaze transition entropy (GTE). SGE provides a measure of overall predictability for fixation locations, which indicates the level of gaze dispersion during a given viewing period. The SGE is calculated using equation (\ref{equation:sge}):
 
\begin{equation}\label{equation:sge}
H(x) = -\sum_{i=1}^{n}(p_i)log_2(p_i) 
\end{equation}

$H(x)$ is the value of SGE for a sequence of data $x$ with length $n$, $i$ is the index for each individual state, $p_i$ is the proportion of each state within $x$. To calculate the SGE, the visual field is divided into spatial bins of discrete state spaces to generate probability distributions. Specifically, the coordinates are divided into spatial bins of 100 × 100 pixel. $i$ to $n$ is defined as all the gaze data during crossing.

GTE is retrieved by applying the conditional entropy equation to first order Markov transitions of fixations with the equation (\ref{equation:gte}):

\begin{equation} \label{equation:gte}
H_{c}(x) = -\sum_{i=1}^{n}(p_i) \sum_{i=1}^{n}p(i,j) log_2 p(i,j) 
\end{equation}

Here $H_{c}(x)$ is the value of GTE, and $p(i, j)$ is the probability of transitioning from state i to state j. The other variables have the same definitions as in the SGE equation (\ref{equation:sge}). More details of calculating SGE and GTE can be found in \cite{Guo2022,guo2022roadway}.

\subsubsection*{Heart Rate}
An Android smartwatch with the “SWEAR” app \cite{boukhechba2020swear} records the HR data with a frequency of 1 Hz. The watch is connected to a smartphone via Bluetooth, and the time is synchronized with the experiment computer before each experiment. All data from the smartwatch is temporally stored on the local device and then uploaded to Amazon S3 cloud storage to download for further analysis.

\subsection*{Participants}
The study is reviewed and approved by the Institutional Review Board for the Social and Behavioral Sciences from University of Virginia (IRB-2148). All experiments were performed in accordance with relevant named guidelines and regulations. All participants are required to sign the consent form prior to the experiment. 51 participants were recruited for the experiment. Most of the participants are local residents, university students, and faculty members who are familiar with the study corridor.
All participants are 18 or older and without color blindness.  For the remaining 49 participants (22 female and 27 male), the mean age is 33.92 with a standard deviation of 12.95 (1 participant did not reveal his/her age information).

\subsection*{Statistical Modeling}
A Linear Mixed Effects Model (LMM) was chosen to model the different response variables between independent variables across participants \cite{brown2021introduction}. The LMM framework is chosen specifically for their ability in addressing random and main effects simultaneously within the same modeling scheme \cite{brown2021introduction}. This type of modeling allows us to investigate the effect of each independent variable by considering that each participant might have different baselines for their psychophysiological responses. For example, participants might have different baselines for heart rate, SGE or head variation. This analysis was performed in R programming language \cite{ihaka1996r} using the LME4 package \cite{bates2007lme4}. All statistical analyses were performed at a 95\% confidence level ($\alpha = 0.05$).

\section*{Data availability statement}
The datasets analysed during the current study are available in the Open Science Framework repository - "Supplemental materials for paper: Rethinking infrastructure design: Evaluating vulnerable road users psychophysiological and behavioral responses to different design alternatives" repository with the link of \url{https://osf.io/8w29f/}.

\bibliography{sample}

\section*{Author contributions statement}
T.D.C., A.H., and A.A. designed the research. X.G., A.A. and E.R. designed the software and conducted the experiment. X.G, A.T. and A.A. analysed the data. X.G, and A.T. wrote the draft. All authors reviewed the manuscript.

\section*{Competing interests}
The authors declare no competing interests.

\end{document}